\documentclass[12pt,a4paper]{article}
\usepackage{graphicx}
\usepackage{dcolumn}
\usepackage{bm}
\usepackage{amsmath,amsfonts,amssymb}
\usepackage{slashed}
\usepackage{braket,xcolor}
\usepackage{verbatim}
\usepackage{subcaption}
\usepackage{multirow}
\usepackage{amsfonts,mathtools,resizegather}
\usepackage[utf8]{inputenc}
\usepackage{caption,jheppub}

\subheader{\begin{flushright}
\texttt{IFT-UAM/CSIC-23-43}
\end{flushright}}

\title{\boldmath Holographic complexity of Jackiw-Teitelboim gravity from Karch-Randall braneworld}
\author[a]{Aranya Bhattacharya,}
\author[b]{Arpan Bhattacharyya,}
\author[c]{and Ayan K. Patra}

\affiliation[a]{Centre for High Energy Physics, Indian Institute of Science,\\ C.V. Raman Avenue, Bangalore 560012, India.}
\affiliation[b]{\textit{Indian Institute of Technology, Gandhinagar, Gujarat-382355, India.}}
\affiliation[c]{Instituto de Fısica Teorica UAM/CSIC, Calle Nicolas Cabrera 13-15, Madrid 28049, Spain.}
\emailAdd{aranyab@iisc.ac.in}
\emailAdd{abhattacharyya@iitgn.ac.in}
\emailAdd{a.patra@csic.es}

\abstract{Recently, it has been argued in \cite{Geng:2022slq} that Jackiw-Teitelboim (JT) gravity can be naturally realized in the Karch-Randall braneworld in $(2+1)$ dimensions. Using the `complexity=volume' proposal, we studied this model and computed the holographic complexity of the JT gravity from the bulk perspective. We find that the complexity grows linearly with boundary time at late times, and the leading order contribution is proportional to the $\varphi_0$, similar to the answer found in \cite{Brown:2018bms}. However, in addition, we find subleading corrections to the complexity solely arising from the fluctuations of these Karch-Randall branes.}

\begin{document}
\maketitle
\flushbottom
\section{Introduction}
The $(1+1)$-dimensional Jackiw-Teitelboim (JT) gravity \cite{TEITELBOIM198341, JACKIW1985343} is one of the simplest models of quantum gravity. This model is handy for studying conjectures relating to the geometry of black holes and scrambling in dual quantum systems \cite{Kitaev, Kitaev1,Almheiri:2014cka, Sachdev:1992fk, Polchinski:2016xgd, Maldacena:2016hyu,Jensen:2016pah, Engelsoy:2016xyb,Sarosi:2017ykf}. Holographically, it produces the nearly conformal dynamics of the $(0+1)$ dimensional Sachdev-Ye-Kitaev (SYK) model at low energies \cite{Maldacena:2016upp}. It also describes the near horizon dynamics of a $(3+1)$ dimensional near-extremal Reissner-Nordstrom (RN) black hole of the Einstein-Maxwell theory upon dimensional reduction \cite{Almheiri:2016fws,Nayak:2018qej, Kunduri:2007vf, Kunduri:2013gce,Castro:2008ms,Bhattacharjee:2020nul}. Moreover, most recently, studies concerning JT gravity have been plenty due to its usefulness in constructing models of black hole evaporation. Especially, JT gravity coupled to the matter has been used in the calculations of quantum extremal surfaces where the island was found, which resulted in the correct Page curve that one expects for a unitary system \cite{Almheiri:2019yqk,Almheiri:2020cfm}. Nevertheless, pure JT gravity without any sort of matter couplings  is also important physically where topological properties of AdS$_2$ manifest \cite{Saad:2019lba,Okuyama:2019xbv,Okuyama:2020ncd}. On the other hand, Karch-Randall (KR) braneworld models \cite{Karch:2000gx,Karch:2000ct} have also played a major role in the research concerning the calculations of the entanglement entropy of backgrounds with and without black holes \cite{Almheiri:2019psy,Geng:2020qvw,Chen:2020uac,Chen:2020hmv,Caceres:2020jcn, Geng:2020fxl,Geng:2021mic,Geng:2021hlu, Caceres:2021fuw,Verheijden:2021yrb}. Loosely speaking, there exist three ways to describe these braneworld models with two branes embedded in any arbitrary dimensions \cite{Geng:2020fxl,Geng:2022tfc}: i) The full \textit{``bulk''} can be described by asymptotically AdS$_{d+1}$ containing two AdS$_d$ branes. ii) The \textit{``intermediate''} description involving both bulk and boundary is understood as the two AdS$_d$ spacetimes (on the two branes) connected through a defect CFT in $(d-1)$ spacetime dimensions at the conformal boundary. iii) Finally, the fully \textit{``boundary''} description is where everything is boiled down to describe properties of the defect CFT$_{d-1}$. Most recently, the KR braneworld model in $(2+1)$ dimensions with two AdS$_2$ branes embedded has been argued to give pure JT gravity in some appropriate limits. Particularly, the authors in \cite{Geng:2022slq, Geng:2022tfc} showed that in $(2+1)$ dimensions, the low energy dynamics of the two fluctuating AdS$_2$ branes, $i.e.$ the intermediate prescription, is described by the pure JT gravity in two particular limits: (a) ``near tensionless limit'' and  (b) ``imposing orbifold symmetry.'' The actions coming from these two limits matches with the action of JT gravity \cite{Geng:2022slq}. The two branes meet at the conformal boundary where the defect CFT description exists. Without any brane fluctuations, the authors argued that the entanglement surface degenerates. From the boundary perspective, this behaviour resonates with the trivial nature of the boundary state where the exact conformal symmetry is respected. However, adding fluctuations to the branes resulted in a unique entangling surface with the same entanglement entropy as that of the JT gravity. In terms of boundary prescription, these fluctuations break down the conformal invariance of the boundary state, thus lifting the degeneracy in the entanglement entropy. Therefore, these small fluctuations are significant for matching the physics of these two setups \cite{Geng:2022slq}. \par

On a different note, holographic complexity has been another way of characterizing quantum gravity states in recent times \cite{susskind2,Susskind:2014rva}. Although the exact definition of quantum complexity is somewhat subtle in the field theory \cite{Jefferson,Chapman:2017rqy,Hackl:2018ptj,Khan:2018rzm,Bhattacharyya:2018bbv,me1,Caputa:2017yrh,Bhattacharyya:2018wym,Caputa:2018kdj,Erdmenger:2020sup,Chagnet:2021uvi, Bhattacharya:2022wlp,Bhattacharyya:2022ren,Bhattacharyya:2023sjr}\footnote{This list is by no means exhaustive. Interested readers are referred to these reviews \cite{Chapman:2021jbh, Bhattacharyya:2021cwf}.}, one can loosely say that quantum complexity tries to estimate the difficulty of preparing a given quantum ``target state", starting with a simple (usually unentangled) ``reference state" using a set of simple universal ``gates" \cite{nielsen2006optimal, Nielsen_2006,NL3}. Holographically, complexity proposals associate either the action or the volume of certain spacetime regions with the quantum complexity of the boundary state. According to the `complexity=volume' proposal, the complexity of a boundary state is given by the volume of the extremal co-dimension one bulk hypersurface anchored at the boundary \cite{Susskind:2014moa},
\begin{equation}
\mathcal{C}_{\text{V}}(\Sigma)=\underset{\Sigma=\partial
\mathcal B}{\text{max}}\left[\frac{\text{Vol}(\mathcal{B})}{G_N \ell}\right]
\label{com=vol}
\end{equation}
where $\mathcal{B}$ is the extremal co-dimension one hyper-surface, $\Sigma$ is the Cauchy slice and $\ell$ is the AdS curvature scale.\footnote{A related conjecture is the subregion complexity proposal \cite{Alishahiha:2015rta,Ben-Ami:2016qex,Agon:2018zso,Hernandez_2021}, which along with the full volume complexity has also been investigated previously in the wedge holography setups \cite{Hernandez_2021,Bhattacharya:2021jrn, Bhattacharya:2021nqj}. Furthermore, a microscopic interpretation of CV in terms of gates on tensor networks was put forward  in \cite{Pedraza:2021fgp,Pedraza:2021mkh,Pedraza:2022dqi}.}
On the other hand, the `complexity=action' proposal dictates that the complexity of the boundary state is equal to the gravitational action evaluated on a particular patch known as Wheeler-de Witt (WDW) patch \cite{Susskind:2014rva, Brown:2015bva},
\begin{equation}
C_{A}(\Sigma)=\frac{\mathcal{I}_{\text{WdW}}}{\pi \hbar}
\label{com=action}
\end{equation}
where $\mathcal{I}_{\text{WdW}}$ is the gravitational action evaluated on the WdW patch. One may think of this WdW patch as the domain of dependence of the maximal volume slice that appears in the CV conjecture.
 
Even though these two conjectures in eqns. \ref{com=vol} and \ref{com=action} do not yield the same results quantitatively for the complexity; they still agree at a qualitative level. The differences in these two bulk quantities are usually thought to be related to the non-uniqueness in the microscopic definition of complexity in the boundary theory, e.g., in the choice of elementary unitary gates \cite{Chapman:2016hwi}. The late-time growth rate of the complexity is proportional to $2M/\pi$, independent of the boundary curvature and the spacetime dimension \cite{Brown:2015bva, Brown:2015lvg}, in CA duality. This late time saturation of the growth rate is related to Lloyd's bound on the rate of computation by a system with energy $M$ \cite{lloyd2000ultimate}. In contrast to the CV proposal, where this late-time growth rate of the complexity also saturates, the final rate is only proportional to the mass at high temperatures and with a coefficient that depends on the spacetime dimension \cite{Stanford:2014jda, Chapman:2016hwi}. From a computational perspective, the complexity is lower bounded by the geodesic distance in a specific manifold \cite{nielsen2006optimal, Nielsen_2006} and it was found that in view of counting the total number of gates required to prepare a unitary operator, complexity naturally scales proportionally to volume after certain optimization \cite{Bhattacharyya:2019kvj}. For CV duality, results from JT and RN agree with each other and match with the expected behaviour of the complexity for SYK quantum mechanics \cite{Brown:2018bms}. Surprisingly, the naive late-time growth rate of the complexity coming from the CA proposal for the JT model vanishes, and this is in conflict with general anticipations for the growth of complexity for chaotic systems like SYK \cite{Brown:2018bms}. However, the late-time complexity growth rate turns out to be non-vanishing only when one treats boundary terms appropriately \cite{Brown:2018bms}. Following the CA conjecture, \cite{Goto:2018iay,Alishahiha:2018swh, Brown:2018bms} discusses this analysis for the complexity growth of the JT gravity. The complexity of JT gravity following CV conjecture and a comparison of that with the Krylov complexity of the dual SYK model has also been discussed in \cite{Jian:2020qpp}. A higher derivative corrected JT-like model has further been discussed in \cite{Banerjee:2021vjy}, which encapsulates the near extremal behaviour of four-dimensional black holes with arbitrary quartic corrections in four dimensions. Furthermore, \cite{Mandal:2022ztj} studied the late-time growth of holographic complexity of a charged black hole in five-dimensional AdS spacetime in the presence of quartic derivative interaction terms using the `complexity = action' conjecture. 

Given the actions of the two theories match exactly $e.g.$, the low energy effective dynamics of the two fluctuating AdS$_2$ branes and the pure JT gravity, one would expect that the action complexity results to match for these two setups in the matching limit. However, it remains an interesting problem to study the complexities of JT gravity and the two KR branes in the braneworld model with fluctuations to check whether or not they agree. Since complexity is conjectured to know more about the evolution of a system than entanglement entropy, it would indeed be a fascinating fact if the complexities of the two cases match exactly as well, strengthening this correspondence at a microscopic level. This would mean that the quantum gravity states dual to the two backgrounds are exactly the same. This motivates us to study the holographic complexity; however, using the complexity equals volume proposal for these two backgrounds and taking the appropriate limit to check this fact explicitly.
More concretely, in this paper, we focus on the late-time complexity growth rate of JT gravity arising from the fluctuations of the KR branes in the appropriate limit. To do so, we will primarily work with a Karch-Randall (KR) brane model (RS braneworlds with sub-critical tension) with two fluctuating AdS$_2$ branes in $(2+1)$ dimensions that were studied in \cite{Geng:2022slq, Geng:2022tfc}\footnote{The authors in \cite{Deng:2022yll} also obtained JT gravity action on the brane by a similar partial dimensional reduction of $(2+1)$ dimensional AdS gravity and the dilaton field in the JT action was found to be related to the fluctuation on the brane.} and also explained in detail in the next section. Our main objective is to understand the effect of these small fluctuations on the holographic complexity of JT gravity.

The rest of the paper is constructed as follows. Section \ref{sec2} is mostly a review part. In section \ref{KRreview}, we review the KR braneworld model, whereas in section \ref{JTreview}, we briefly note down the appropriate details for JT gravity. Then we discuss the holographic complexity of JT gravity in section \ref{JTCV}. In section \ref{JTfromKR}, we review how at the level of the action the two backgrounds can be mapped, provided one considers fluctuating branes in the braneworld model. In Section \ref{QImatching}, we first explain the matching of the entanglement entropy for the two cases (section \ref{EEJTKR}). Then we \ compare the computations for volume complexity of JT gravity (section \ref{JTCV}) and the braneworld model with fluctuating branes (section \ref{mainresult}). Finally, in section \ref{conc}, we conclude with the main results, explanations and future directions.

\section{Basic review}\label{sec2}
\subsection{Karch-Randall Braneworld Models}\label{KRreview}
We start this section by briefly reviewing the KR braneworld model \cite{Karch:2000gx,Karch:2000ct} in $(d+1)$ dimensions with single and double branes embedded. The following action describes the braneworld model with a single brane embedded,
\begin{equation}
\mathcal{S}=-\frac{1}{16\pi G_{d+1}}\int d^{d+1}y\sqrt{-g}(R-2\Lambda)-\frac{1}{8\pi G_{d}}\int d^{d}y\sqrt{-h}(K-T)
\label{1st}
\end{equation}
where $R$ is the Ricci scalar of the bulk space-time and $\Lambda=-\frac{d(d-1)}{2\ell^2}$, is the cosmological constant; $K$ is the trace of extrinsic curvature of the embedded brane with induced metric $h_{ab}$ and $T$ represents the brane tension with the condition $T\leq (d-1)$. We impose the Neumann boundary condition on the metric fluctuations near the brane \cite{Geng:2022slq,Geng:2022tfc},
\begin{equation}
    \nabla_{n}\partial g_{\alpha\beta}\big|_{\text{near brane}}=0
\end{equation}
where $n$ represents the normal direction to the brane. Variation of the action \ref{1st} with respect to bulk metric $g$ leads to standard Einstein's equation with a negative cosmological constant $\Lambda$,
\begin{equation}
    R_{\alpha \beta}-g_{\alpha\beta}+\Lambda g_{\alpha\beta}=0\label{eneq}
\end{equation}
On the other hand, the brane embedding is determined by the Israel junctions condition,
\begin{equation}
    K_{ab}=(K-T)h_{ab}\label{isjun1}
\end{equation}

It is straightforward to check that AdS$_{d+1}$ black string geometry satisfies the equation of motion in \ref{eneq} with the metric,
\begin{equation}
ds_{d+1}^2=d\rho^2+\ell^2\cosh^2(\rho/\ell) ds^2_{\text{AdS}_{\text{BH}_d}}
\label{2nd}
\end{equation}
where $\rho$ is the radial direction ranges from $-\infty\,\,\text{to}\,\,+\infty$. The brane is located at $\rho=\text{constant}$ and essentially describes the planar AdS$_d$ black hole geometry. The metric on the brane is given by,
\begin{equation}
  ds^2_{\text{AdS}_{\text{BH}_d}}=\frac{1}{u^2}\bigg(-f(u)d\tilde{t}^2+\frac{du^2}{f(u)}+d\vec{x}^2_{d-2}\bigg)  
\end{equation}
where $u$ is the radial direction on the brane with a horizon at $u=u_h\,\,\text{and}\,\,f(u)=\big(1-\frac{u^{d-1}}{u_h^{d-1}}\big)$, is the blackening factor. One can think of this bulk AdS$_{d+1}$ black string geometry as a foliation of planar AdS$_d$ black holes at each constant radial slice $i.e.$ $\rho=\text{constant}$. We excise the bulk region beyond the brane location $\rho=\rho_b$ to the conformal boundary.
In the language of ``double holography''\cite{Almheiri:2019psy,Almheiri:2020cfm}, this system has three equivalent descriptions:
\begin{itemize}
\item[(I)] a $d$-dimensional boundary conformal field theory (BCFT), i.e. a $d$-dimensional CFT with a $(d-1)$-dimensional boundary \cite{MCAVITY1995522,Cardy:2004hm},
\item[(II)] a $d$-dimensional CFT coupled to gravity on an asymptotically AdS$_d$ space $\mathcal{M}_d$, with a half-space CFT bath coupled to $\mathcal{M}_d$ via transparent boundary conditions at an interface point,
\item[(III)] Einstein gravity on an asymptotically AdS$_{d+1}$ space containing $\mathcal{M}_d$ as an ``end-of-the-world" brane \cite{PhysRevLett.83.4690,Karch:2000ct,Karch:2000gx}.
\end{itemize}
Note that the scenario relevant for the black hole information paradox is (II) \footnote{For a comprehensive review, interested readers are referred to \cite{Almheiri:2020cfm,Raju:2020smc}. Also, recently, the author in \cite{Yadav:2023qfg} applied wedge holography to study different entanglement properties of the multiverse.}. Prescription (I) and (III) are related $via$ \textit{AdS/BCFT correspondence} \cite{Takayanagi:2011zk,Fujita:2011fp} whereas (I) and (II) is related $via$ \textit{AdS/CFT correspondence}. The advantage of such doubly holographic models is that the interesting semiclassical physics arising from (II) can be extracted using computations performed classically using prescription (III) \cite{Almheiri:2019hni,Akal:2020twv, Akal:2021foz,Chen:2020uac}. More precisely, the generalized entropy of (II) is well-approximated, to leading order in $1/G_N$, in (III) by a classical entanglement surface computed via the \textit{Ryu-Takayanagi (RT)} prescription \cite{Ryu:2006bv} (or its covariant extension \cite{Hubeny:2007xt})---the surface is extremal and thus must satisfy appropriate boundary condition on the brane. Hence, this equivalence allows us to interpret the quantum extremal surfaces (QES) \cite{Engelhardt:2014gca,Wall:2012uf} in prescription (II) as RT/HRT surfaces in prescription (III). Technically speaking, it is the matter entropy $S_{\text{matter}}$, which is well-approximated by such an area. However, so long as the only gravitational terms on the brane are “induced” by gravity in the bulk, the $G^{-1}_{d}$ term vanishes at tree level and thus counts as a quantum correction which we neglect in a semiclassical approximation taking only an effective theory on the brane. \cite{Chen:2020uac, Geng:2020fxl} disscuss this in detail. Nevertheless, a particularly useful manifestation of double holography is when the end-of-the-world brane is ``tensionless" in the sense of the Karch-Randall-Sundrum constructions \cite{Randall:1999vf,Karch:2000ct,Karch:2000gx}. While such a ``probe" brane does not backreact on the bulk geometry of (III), there is still a tower of spin-2 Kaluza-Klein (KK) modes living on the brane \cite{Karch:2000ct}. As discussed in \cite{Geng:2020qvw}, one may still consider scenario (II) by taking the lowest-mass mode as a graviton and the higher modes to compose the CFT. While such a theory is not standard Einstein gravity, the result of using a tensionless braneworld is that holographic calculations in bulk (III) do not require particularly intricate numerics, unlike in setups with nontrivial tension parameters \cite{Almheiri:2019psy,Geng:2020fxl}.
\begin{figure}
    \centering
    \includegraphics[scale=0.4]{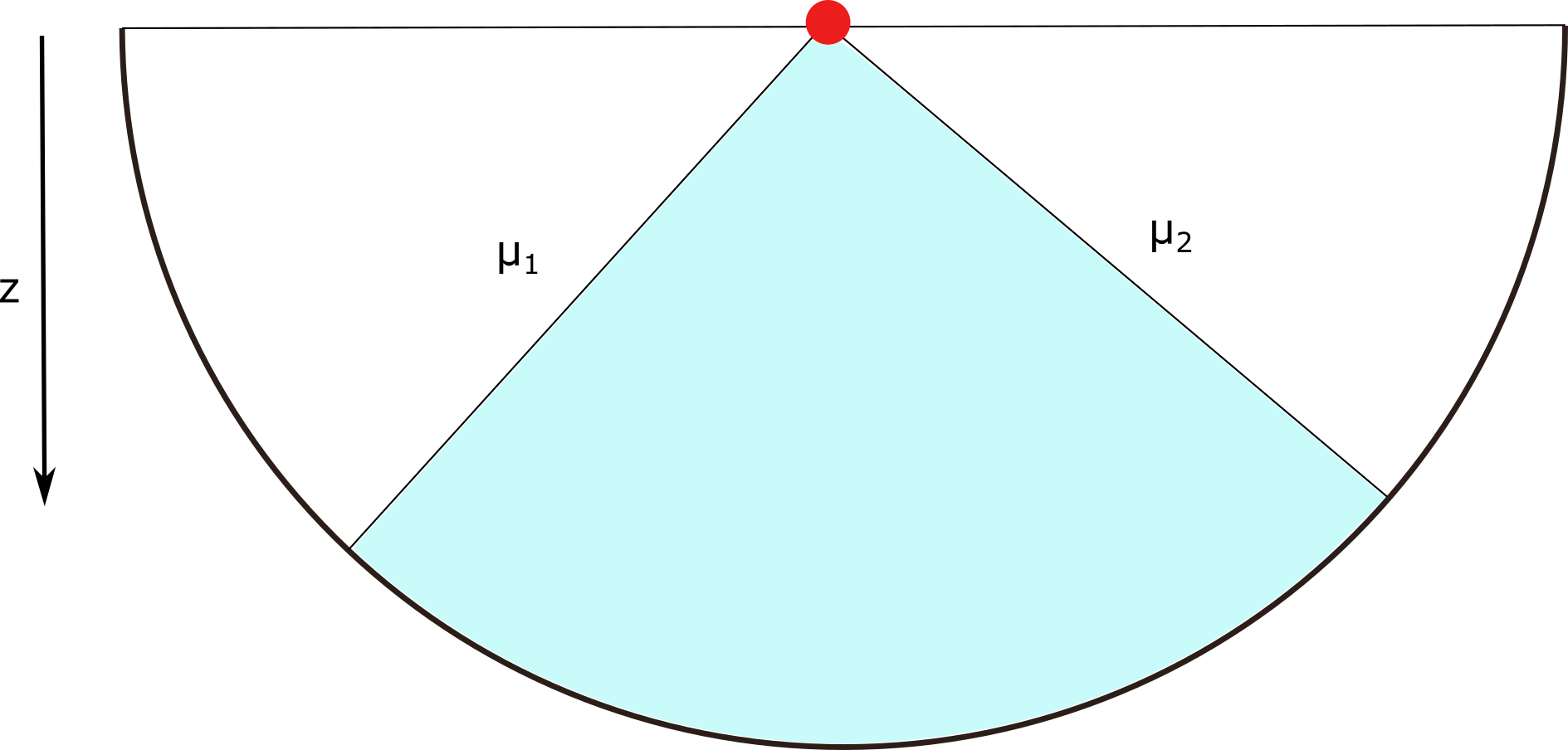}
    \caption{Two rigid branes embedded in bulk AdS space-time at the locations $\mu=\mu_1$ and $\mu=\mu_2$. The angular coordinate $\mu$ is related to the bulk radial direction $\rho$ \textit{via} a coordinate transformation, $\rho=\log{\cot{\frac{\mu}{2}}}$. The two branes meet at the conformal defect represented by the red dot. We excise the region beyond the brane location to the boundary. This results in the left-over bulk region shaded in light blue. The volume of this light blue region represents the complexity of the boundary state.}
    \label{rigidbranes}
\end{figure}
\par
Let us now discuss the double KR braneworld model. In this setup, we add another positive tension brane which is described by the action,
\begin{equation}
\mathcal{S}=-\frac{1}{16\pi G_{d+1}}\int d^{d+1}y\sqrt{-g}(R-2\Lambda)-\frac{1}{8\pi G_{d}}\int_{j=1,2} d^{d}y\sqrt{-h}(K_j-T_j)
\label{3rd}
\end{equation}
The bulk geometry is still given by  \ref{2nd} and we have two branes localized at $\rho=\rho_1$ and $\rho=\rho_2$ with respective tensions $T_1=(d-1)\tanh(-\rho_1/\ell)$ and $T_2=(d-1)\tanh(\rho_2/\ell)$ with $\rho_1<0$ and $\rho_2>0$.
Similar to the single brane scenario, we excise the bulk region that is beyond the brane location $i.e.$, $\rho=\rho_1\,\text{and}\,\rho=\rho_2$. In other words, we only consider the bulk region that is enclosed by these two branes. In the language of ``wedge holography'', it has three equivalent prescriptions \cite{Geng:2020fxl,Geng:2022tfc,Geng:2022slq,Akal:2020wfl}:
\begin{itemize}
    \item[(I)] a $(d-1)$-dimensional conformal field theory (CFT);
    \item[(II)]two $d-$dimensional CFTs coupled to gravity on asymptotically AdS$_d$ spaces $\mathcal{M}^{(1)}_{d}$ and $\mathcal{M}^{(2)}_{d}$, with these two systems connected via a transparent boundary condition at the $(d-1)$ dimensional defect;
    \item[(III)] Einstein gravity on an AdS$_{d+1}$ space containing two AdS$_d$ branes $\mathcal{M}^{(1)}_{d}$ and $\mathcal{M}^{(2)}_{d}$, which intersect each other on the asymptotic boundary thus forming a wedge;
\end{itemize}
The scenario that is relevant for the black hole information paradox is (II), similar to the single brane case\cite{Geng:2020fxl}. However, unlike the single-brane situation where only one brane is gravitating, here, both of the branes are gravitating; thus, one can consider a situation where the thermal bath is also gravitating in the models of black hole evaporation. This consideration results in a constant Page curve in higher dimensions ($d>2$) \cite{Geng:2020fxl}. However, The exact Page curve appears only when one bipartite the whole system across the defect and considers entanglement between these two sub-systems. Strikingly, in the $(2+1)$ dimension, this system gives rise to interesting physics when one considers fluctuating branes in contrast to its higher dimensional counterpart, where the brane fluctuations can be ignored \cite{Geng:2022slq}. For a better understanding of this model, we will study this double-brane system in $(2+1)$ dimensions and try to compute the total complexity of the corresponding microstates.
\subsection{Jackiw-Teitelboim gravity}\label{JTreview}
In this section, we briefly review pure Jackiw-Teitelboim gravity (JT) gravity. This is discussed in \cite{Almheiri:2019psf} in the context of quantum extremal surfaces. Additionally, it is one of the descriptions for the doubly holographic model of \cite{Almheiri:2019hni} when coupled to conformal matter. We use the action in \cite{Harlow:2018tqv}.\footnote{The action presented by \cite{Harlow:2018tqv} includes a holographic renormalization meant to keep the action finite on relevant classical configurations.} In particular, we show that the action is extremized for configurations involving fixed AdS$_2$ backgrounds, with the matter stress tensor being related to the dilaton by an additional set of on-shell constraints. We also specify boundary conditions on the metric and the dilaton. The action of JT gravity itself consists of two separate parts,
\begin{equation}
I_{JT}[g_{ij}^{(2)},\varphi] = I_{T}[g_{ij}^{(2)}] + I_{G}[g_{ij}^{(2)},\varphi],
\end{equation}
where these terms are defined as,
\begin{align}
I_{T}[g_{ij}^{(2)}] &= \frac{\varphi_0}{16\pi G_2} \left(\int_{\mathcal{M}} d^2 x\sqrt{-g}R + 2\int_{\partial\mathcal{M}} dx \sqrt{|\gamma|}K\right),\\
I_{G}[g_{ij}^{(2)},\varphi] &= \frac{1}{16\pi G_2}\left[\int_{\mathcal{M}} d^2 x \sqrt{-g}\varphi\left(R + \frac{2}{\ell^2}\right) + 2\int_{\partial\mathcal{M}} dx \sqrt{|\gamma|}\varphi(K-1)\right].\label{dynamicalAct}
\end{align}
Here, $g_{ij}^{(2)}$ is the background metric and $\varphi$ is the dynamical dilaton. We also have couplings $G_2$ and $\varphi_0 \gg \varphi$. Note that $I_G$ is a dynamical term, whereas $I_{T}$ is topological. This becomes obvious by the Gauss-Bonnet theorem in the Euclidean sector; for a 2-dimensional orientable, Riemannian manifold $\mathcal{M}_E$ with Euler characteristic $\chi(\mathcal{M}_E)$,
\begin{equation}
\int_{\mathcal{M}_E} d^2 x\sqrt{g_E}R + 2\int_{\partial\mathcal{M}_E} dx \sqrt{\gamma_E}K = 4\pi\chi(\mathcal{M}_E).
\end{equation}

Thus, the Euclideanized $I_T$ in the JT gravity action is,
\begin{equation}
I^E_{T} = -\frac{\varphi_0 \chi(\mathcal{M}_R)}{4G_2} = \frac{\varphi_0}{4G_{2}}(2g + b - 2),
\end{equation}
where $g$ is genus and $b$ is the number of boundaries.
In the path integral, any term of the form $\exp(-I_T^E)$ corresponding to a configuration with large $g$ or large $b$ will be exponentially suppressed. Consequently, in the approximation for which we consider the leading-order term, we take $g = 0$ and $b = 1$. As mentioned in \cite{Almheiri:2019psf}, this means that, semiclassically, the topological term yields the following leading-order contribution to the entropy,
\begin{equation}
S_T \approx \log \exp\bigg(\frac{\varphi_0}{4G_2}\bigg) = \frac{\varphi_0}{4G_2}.
\end{equation}

Now, in finding the classical configurations for the 2-dimensional bulk, we can neglect the variation of $I_T$ (since it is topological). Furthermore, we apply the Dirichlet boundary conditions in \cite{Harlow:2018tqv,Almheiri:2019psf} to fix the boundary metric and the boundary value of the dilaton,
\begin{equation}
\gamma_{uu}|_{\partial\mathcal{M}} = \frac{1}{\epsilon^2},\ \ \varphi|_{\partial\mathcal{M}} = \frac{\varphi_b}{\epsilon},\label{dilatonBC}
\end{equation}
taking $\epsilon \to 0$ and $\varphi_b > 0$ finite. With these boundary conditions, we can neglect the variation of the boundary terms in the action. We now focus on varying just the bulk parts of $I_G$ and $I_{W}$. First, varying by $\varphi$, we find that,
\begin{equation}
\frac{\delta}{\delta \varphi}I_{G} = \frac{1}{16\pi G_2} \sqrt{-g}\left(R + \frac{2}{\ell^2}\right).\label{eomBackground}
\end{equation}

Classically, we thus have that the scalar curvature is fixed. Furthermore, this extends to the semiclassical regime because we do away with any quantum backreaction in the metric. Specifically, the background must be locally AdS$_2$,
\begin{equation}
R = -\frac{2}{\ell^2}.\label{eomBackground2}
\end{equation}

Upon computing the variation of \eqref{dynamicalAct}, we find that,
\begin{align}
\frac{\delta}{\delta g^{ij}}I_{G}
=\ &\frac{\sqrt{-g}}{16\pi G_2}\left[-\frac{1}{2}g_{ij}\left(R + \frac{2}{\ell^2}\right)\varphi + R_{ij}\varphi - \nabla_i \nabla_j \varphi + g_{ij}\nabla^2 \varphi\right]\nonumber.
\end{align}

We can apply \eqref{eomBackground2} to eliminate the first term in the brackets. Additionally, setting this to $0$ yields the following classical equations of the motion for the dilaton,
\begin{equation}
(R_{ij} + g_{ij}\nabla^2 - \nabla_i \nabla_j)\varphi=0.\label{stress1}
\end{equation}
Interestingly, when JT gravity is classically coupled to a CFT,  the matter stress tensor is coupled to the dilaton, while the background has a fixed Ricci curvature. Since we neglect any quantum backreaction on the metric in the semiclassical approximation, the background remains fixed, and occurrences of the stress tensor are replaced with its expectation value \cite{Almheiri:2019psf}. Nevertheless, we consider a local AdS$_2$ patch of the background in lightcone coordinates,
\begin{equation}
\frac{ds^2}{\ell^2} = -\frac{4dx^+ dx^-}{(x^+ - x^-)^2}.
\end{equation}
In these coordinates, the three independent components of \eqref{stress1} are,
\begin{align}
0 &= \partial_+ \partial_- \varphi + \frac{2}{(x^+ - x^-)^2}\varphi,\\
0 &= -\frac{1}{(x^+ - x^-)^2}\partial_+ \left[(x^+ - x^-)^2 \partial_+ \varphi\right],\\
0 &= -\frac{1}{(x^+ - x^-)^2}\partial_- \left[(x^+ - x^-)^2 \partial_- \varphi\right].
\end{align}
The solution for the dilaton in this coordinate is given by \cite{Almheiri:2019psf},
\begin{equation}
    \varphi(x^+,x^-)=\frac{\varphi_h}{\sqrt{\mu}}\frac{1-\mu x^+ x^-}{x^+-x^-}
\end{equation}
where $\mu$ and $\varphi_h$ are real parameters. We perform a further coordinate transformation which maps the light cone coordinates to Schwarzschild-like geometry in $(1+1)$ dimension,
\begin{eqnarray}
    r&=&\frac{1-\mu x^{+}x^{-}}{x^{+}-x^{-}}\\
    t&=&\frac{1}{\sqrt{\mu}}\text{arccoth}\bigg({\frac{1}{\sqrt{\mu}}\frac{1+\mu x^{+}x^{-}}{x^{+}+x^{-}}}\bigg)
\end{eqnarray}
This transformation yields,
\begin{equation}
    ds^2=-(r^2-r^2_h)dt^2+\frac{dr^2}{r^2-r^2_h};\,\,\,\text{with}\,\,\mu=r^2_h.
\end{equation}
The dilaton then becomes a function of only the radial direction $r$; \textit{i.e.} $\varphi=\frac{\varphi_h}{r_h}r$.
\subsection{Holographic complexity of JT gravity}\label{JTCV}

This section reviews the JT gravity complexity, primarily following \cite{Brown:2018bms}. We will use the `complexity=volume' proposal to compute the holographic complexity of JT gravity. For that purpose,
we need to find the maximal volume slice for the background metric of the JT gravity. We choose the Schwarzschild-like $(1+1)$ dimensional black hole geometry as a background for the JT gravity, which is,
\begin{equation}
    ds^2=\frac{L^2}{z^2}\bigg[-h(z)dt^2+\frac{dz^2}{h(z)}\bigg]
\end{equation}
where  blackening factor $h(z)$ and the dilation profile $\varphi(z)$ is given by,
\begin{equation}
    h(z)=\bigg(1-\frac{z^2}{z_h^2}\bigg),\,\,\,\,\,\,\,\,\varphi(z)=\varphi_h\frac{z_h}{z}
\end{equation}\\
and $L$ is the AdS curvature scale.
We parametrized the volume slice as $t\equiv t(z)$. To find the maximal volume slice, we need to extremize the following volume functional,
\begin{equation}
    \mathcal{V}=L \int \frac{dz}{z}\sqrt{-h(z)t'(z)^2+\frac{1}{h(z)}}\equiv L\int dz \mathcal{L}\label{volJT}
\end{equation}
In the above expression for the volume functional, the Lagrangian $\mathcal{L}$ does not explicitly depends on $t$; thus, we can determine the conserved quantity $\mathcal{E}$ from the Lagrangian which is constant on the entire hyper-surface,
\begin{equation}
    \mathcal{E}=-\frac{\partial\mathcal{L}}{\partial {t'(z)}}\implies t_b=-\int_{0}^{z_t}\frac{\mathcal{E} z dz}{h(z) \sqrt{\mathcal{E}^2 z^2+h(z)}}\label{eqnconserveJT}
\end{equation}
where $z_t$ is the turning point for the symmetric geodesic. At this point, the time derivative of the radial direction vanishes $i.e.$, $\frac{1}{t'(z_t)}=0\implies z_t= \frac{z_h}{\sqrt{1-\mathcal{E}^2z_h^2}}$.\\
Putting all these things together, we find the maximal volume after replacing \ref{eqnconserveJT} into \ref{volJT}. 
\begin{figure}
    \centering
    \includegraphics[scale=0.8]{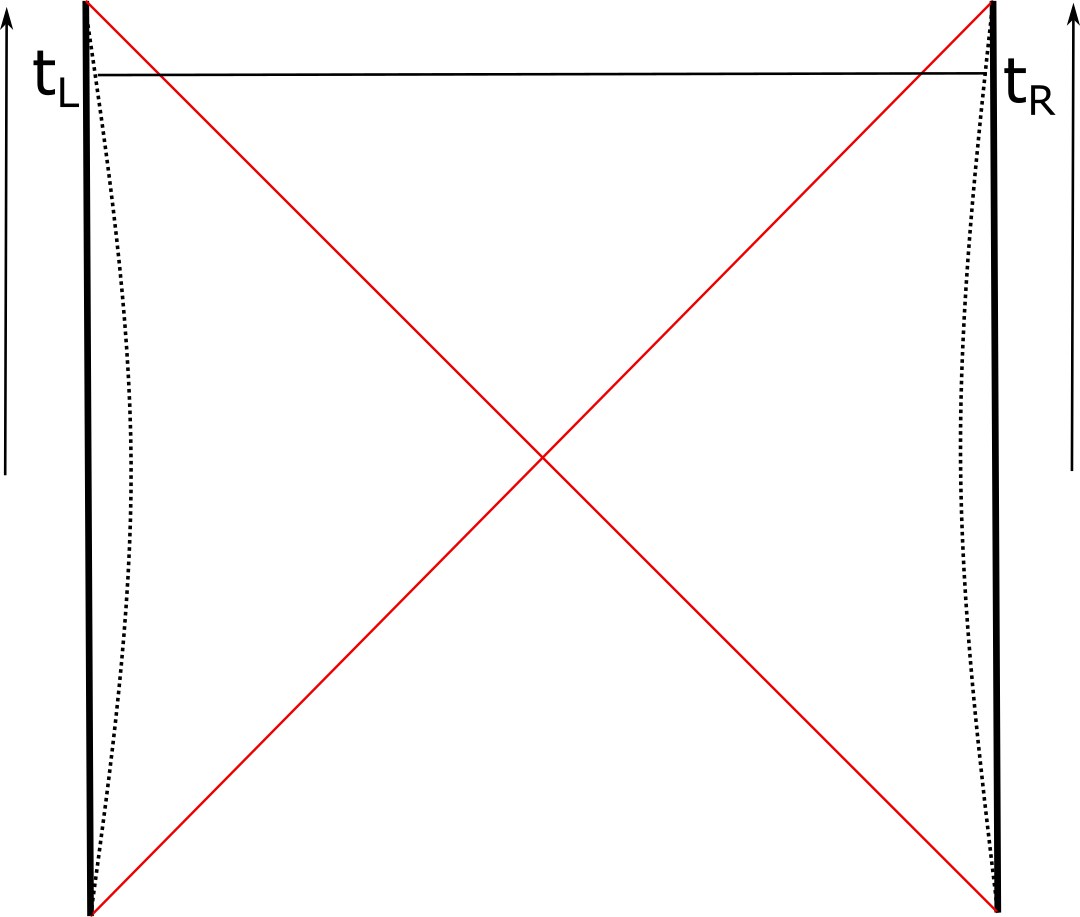}
    \caption{ The horizontal black line represents the maximal volume slice in JT gravity at the boundary time $t_b=t_L=t_R$. }
    \label{fluctuatingbranes}
\end{figure}
This yields the expression for the maximal volume,
\begin{equation}
    \mathcal{V}=L \int_{0}^{z_t}\frac{dz}{z\sqrt{\mathcal{E}^2 z^2+h(z)}}\label{maxvolJT}
\end{equation}
Note that as the boundary time, $t_b$ approaches $\infty$, the conserved quantity $\mathcal{E}$ approaches to some critical value $\mathcal{E}_{\text{crit}}$. This critical value is determined by extremizing $\mathcal{E}$ as the function of turning points $z_t$. This extremization gives,
\begin{equation}
    \mathcal{E_\text{crit}}=\frac{1}{z_h}
\end{equation}
In other words, as we go closer and closer to the infinite boundary time, the energy for these spacelike surfaces approaches some fixed constant value. After substituting $\mathcal{E}_{\text{crit}}$ into \ref{maxvolJT} for the late times and taking the time derivative, we find the late time complexity growth \footnote{We use the relation $G_3=L G_2$.},
\begin{equation}
    \frac{d\mathcal{C}_{\text{JT}}}{d t_b}\bigg|_{t_b\to \infty}=\frac{2\varphi_0}{G_3 z_h}\label{comJT}
\end{equation}

Following the argument of \cite{Brown:2018bms}, we put the $\varphi_0/L$ factor by hand. This is well justified by the expectation that the complexity should grow at a rate proportional to the number of degrees of freedom of the dual quantum system. The number of degrees of freedom is, in turn, proportional to the black hole entropy, and for the $(1+1)$-dimensional black holes, the entropy is dominated by the extremal entropy, $S_0=\frac{\varphi_0}{4 G_2}$\cite{Brown:2018bms}.
Surprisingly, we will see later that this type of factor naturally occurs when one computes the complexity of JT gravity from the Karch-Randall braneworlds with fluctuating branes.

\subsection{JT gravity from the KR braneworld}\label{JTfromKR}
In this section, we briefly review the emergence of JT gravity as a low energy effective dynamics of fluctuating AdS$_2$ branes in bulk AdS$_3$. Before that, note that in the single brane KR setup, the location of the brane is fixed. In that case, One can take the brane location to be an orbifold fixed point to model an End-of-the-world (EOW) brane analogues to $O$- planes in string theory \cite{polchinski_1998, johnson_2002}. However, with two branes, the relative distance between two branes can vary and thus can be treated as a dynamical variable known as radion \cite{Arkani-Hamed:2000ijo}. To study the low-energy effective theory, we will consider only small fluctuations of these AdS$_2$ branes about their rigid locations in $(2+1)$ dimensional bulk AdS. More concretely, let us consider two AdS$_2$ branes located at $\rho=\rho_1+\delta\varphi_1(y)$ and $\rho=\rho_2+\delta\varphi_2(y)$ with the tensions $T_1$ and $T_2$ respectively, where the notation $y$ symbolise orthogonal coordinates. These branes meet at the conformal boundary and thus form a wedge. In addition, these brane fluctuations are considered to be small with respect to the AdS curvature scale \textit{i.e.} $\delta \varphi_1/\ell\ll 1$ \& $\delta \varphi_2/\ell\ll 1$. After putting all these together, the bulk metric can be written as,
\begin{eqnarray}
ds_b^2&=&d\rho^2+\ell^2\cosh^2(\rho/\ell) g_{ab}(\rho, y)dy^a dy^b\\
&\approx& d\rho^2+\ell^2\cosh^2(\rho/\ell) g_{ab}(y)dy^a dy^b
\label{4th}
\end{eqnarray}
The above metric describes the lowest 2d graviton mode as we can disregard higher order terms in $g_{ab}(\rho,y)$ as $\delta \varphi_1/\ell, \delta \varphi_2/\ell\ll 1$ so, $g_{ab}(\rho,y)\approx g_{ab}(0,y)\equiv g_{ab}(y)$. We now plug \ref{4th} back into equation \ref{3rd} and find the leading order contribution in the action up to $\mathcal{O}(\delta\varphi^2)$,
\begin{equation}
\mathcal{S}_{\text{eff}}=\mathcal{S}_0-\frac{1}{16\pi G_3}\int d^2y \sqrt{-g}\varphi(y)(R+2/\ell^2)+\mathcal{S}_{\text{dilaton}}\label{JTeff}
\end{equation} 
where we have introduced a new variable, $\varphi(y)=\delta\varphi_2(y)-\delta\varphi_1(y)$ and $\mathcal{S}_0$ is a completely topological term given by \cite{Geng:2022slq,Geng:2022tfc},
\begin{eqnarray}
\mathcal{S}_0&=&-\frac{(\rho_2-\rho_1)}{16\pi G_3}\int d^2x \sqrt{-g}R[g]=-\frac{\varphi_0}{16\pi G_3}\int d^2x \sqrt{-g}R[g]\\
\mathcal{S}_{\text{dilaton}}&=&-\frac{1}{8\pi G_3}\int d^2 y\sqrt{g}\bigg[\frac{\tanh{\rho_2}}{2}\nabla_\alpha\delta\varphi_2\nabla^{\alpha}\delta\varphi_2+\tanh{\rho_1}(\delta\varphi_1)^2\\
&-&\frac{\tanh{\rho_1}}{2}\nabla_\alpha\delta\varphi_1\nabla^{\alpha}\delta\varphi_1-\tanh{\rho_1}(\delta\varphi_1)^2\bigg]
\end{eqnarray}\\
where $\varphi_0$ is defined in terms of the fixed distance between the two branes $, i.e., \varphi_0=(\rho_2-\rho_1)$.
We can neglect $S_{\text{dilaton}}$ in two cases: $(\textit{i})$ in the near tensionless limit when $T_1, T_2\to 0$\footnote{Note that when $T_1=T_2=0$, the topological term exactly vanishes, as well as $\mathcal{S}_{\text{dilaton}}$. However, the dynamical term still exists for the exactly tensionless branes in \ref{JTeff}. We work in a slightly different limit when $T_1, T_2 \ll 1$ and the brane fluctuations are much smaller than the brane locations. In this limit, the topological term in the pure JT gravity exists, and the leading order term in $\mathcal{S}_{\text{dilaton}}$ vanishes.}, or $(\textit{ii})$ by imposing orbifold symmetry, $\delta\varphi_2(y)=-\delta\varphi_1(y)$ \cite{Geng:2022slq, Geng:2022tfc}.
Both of these cases lead to the bulk action up to $\mathcal{O}(\varphi^2)$,
\begin{equation}
\mathcal{S}_{\text{eff}}=\mathcal{S}_0-\frac{1}{16\pi G_3}\int d^2y \sqrt{-g}\varphi(y)(R+2/\ell^2)
\end{equation} 
\begin{figure}
    \centering
    \includegraphics[scale=0.5]{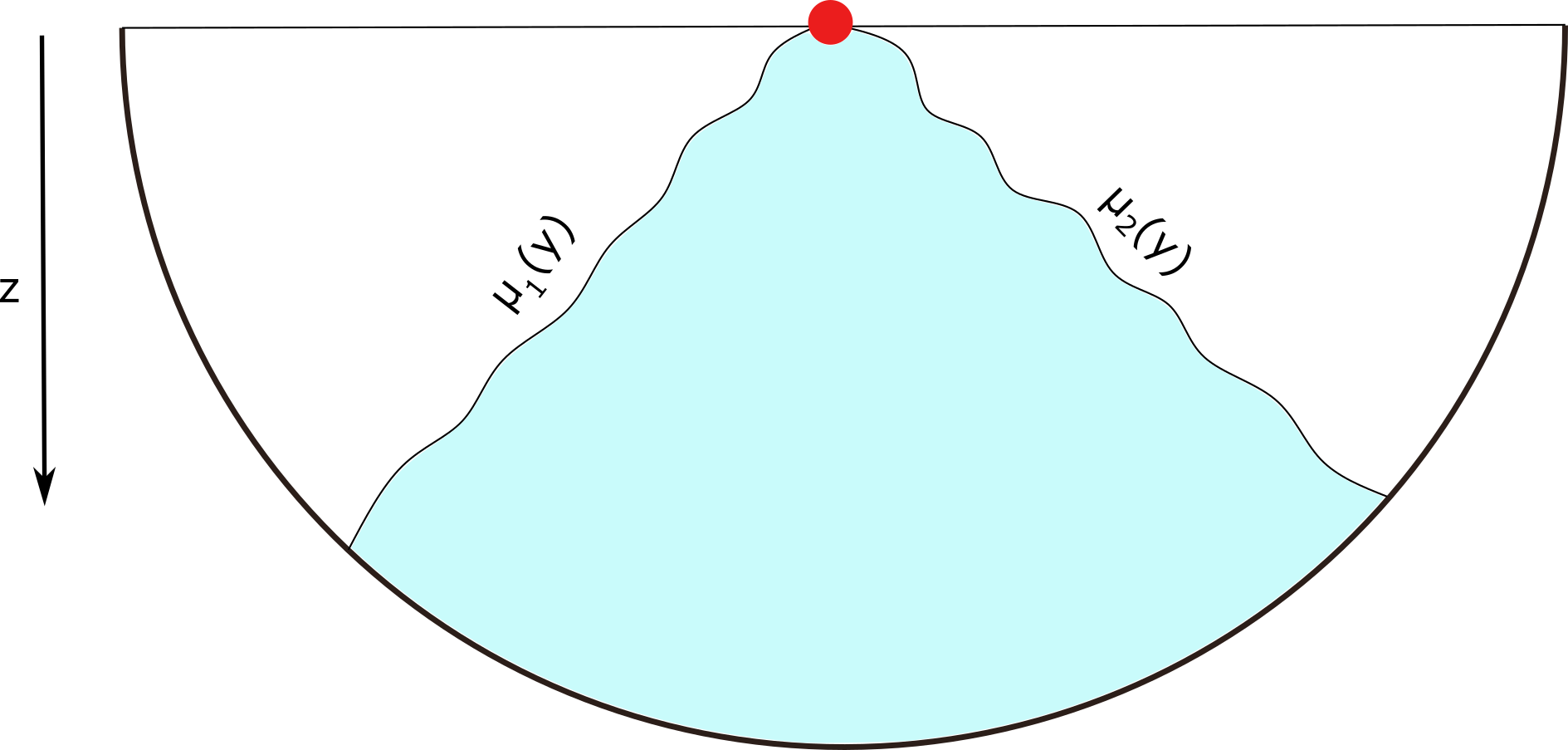}
    \caption{Two fluctuating branes embedded in bulk AdS space-time. The brane location is now a function of the orthogonal directions represented by collective notation $y$. We excise the region beyond the brane position to the conformal boundary and consider the bulk portion shaded in light blue. The low energy effective dynamics of the fluctuating branes are then correctly captured by the JT gravity.}
    \label{fluctuatingbranes}
\end{figure}
This is exactly the pure JT gravity action. To study non-trivial dynamics, one further imposes boundary conditions and a cutoff on the bulk metric,
\begin{equation}
    g_{ab}\big{|}_{\text{bdy}}=-\frac{1}{\epsilon^2},\,\,\,\,\varphi(y)\big{|}_{\text{bdy}}=\frac{\varphi_b}{\epsilon}.
\end{equation}
where $\varphi_b=\varphi_h z_h$.
With this identification of boundary condition and cutoff for the bulk AdS$_3$ metric, we must add the Gibbons Haling boundary to the total action,
\begin{equation}
    \mathcal{S}_{\text{GHY}}=-\frac{1}{8\pi G_3}\int d^2 y \sqrt{-h^{\text{bdy}}}K^{(3)},
\end{equation}
where $h_{ab}^{\text{bdy}}$ is the induced metric on the cutoff surface, and $K^{(3)}$ is the trace of the extrinsic curvature of this surface embedded in the 3$d$ bulk. This term precisely leads to the boundary term in the JT gravity,
\begin{equation}
    \mathcal{S}_{\text{bdy}}=-\frac{2\varphi_0}{16\pi G_3}\int_{\partial\mathcal{M}}dy \sqrt{-h}K-\frac{2\varphi_b}{16\pi G_3}\int_{\partial\mathcal{M}}dy\sqrt{-h} K
\end{equation}
where K is the trace of the extrinsic curvature of the cutoff boundary of the AdS$_2$.
With this boundary term added, the full action of low energy theory arising from brane fluctuation is \cite{Geng:2022slq, Geng:2022tfc},
\begin{eqnarray}
    \mathcal{S}_{\text{eff}}&=&-\frac{\varphi_0}{16\pi G_3}\bigg[\int_{\mathcal{M}}d^2y\sqrt{-g}R[g]+2\int_{\partial\mathcal{M}}dy \sqrt{-h}K\bigg]\\
                            & &-\frac{1}{16\pi G_3}\bigg[\int_{\mathcal{M}}d^2y\sqrt{-g}\varphi(y)(R[g]+2/\ell^2)+2\varphi_b\int_{\partial\mathcal{M}}dy \sqrt{-h}K\bigg]
\end{eqnarray}\\
By varying the above action with respect to the metric tensor and the dilaton field, we arrive at the classical equation of motion,
\begin{eqnarray}
    0&=&R+2/\ell^2\\
    0&=&(R_{ij} + g_{ij}\nabla^2 - \nabla_i \nabla_j)\varphi.
    \label{eineq}
\end{eqnarray}
The fixed AdS$_2$ black hole metric can be considered as the background metric for the JT gravity,
\begin{eqnarray}
    ds^2&=&\frac{1}{z^2}\bigg[-h(z)dt^2+\frac{dz^2}{h(z)}\bigg]
\end{eqnarray}
where $h(z)=\big(1-\frac{z^2}{z_h^2}\big)$, is the blackening factor.
By solving Einstein's equation \ref{eineq}, one finds the following profile for the dilaton field,
\begin{equation}
    \varphi(z, t)=\varphi_h\frac{z_h}{z}.
\end{equation}

\section{Matching of entanglement and complexity}\label{QImatching}

\subsection{Entanglement Between Two Defects}\label{EEJTKR}
In this section, we compute the entanglement entropy for the thermal state in $(2+1)$ dimension. We consider the nearly tensionless limit and find the minimal length connecting the two branes through the bulk AdS. When the branes are rigid, the minimal surface lies on the constant time slice. Thus we need to extremize the following length functional,
\begin{equation}
\mathcal{A}=\int d\rho\sqrt{1+\frac{\ell^2\cosh^2{(\rho/\ell)}}{z^2 h(z)}z'(\rho)^2}\label{rigidlength}
\end{equation}
By making the coordinate transformation $z\to z^*(z)$, \ref{rigidlength} become,
\begin{equation}
  \mathcal{A}=\int d\rho\sqrt{1+{\ell^2\cosh^2{(\rho/\ell)}}{z^{*}}'(\rho)^2},\,\,\,\,\,\,\,\,\,\,\,z^*=\int\frac{dz}{z\sqrt{h(z)}}\label{rigidlength1}  
\end{equation}\\
Upon extremization \ref{rigidlength1} leads to ${z^{*}}'(\rho)=0$. Thus any $z^*=\text{constant}$ slices represent an equal area or entanglement entropy between two defects,
\begin{equation}
    S_{\text{EE}}=\frac{(\rho_2-\rho_1)}{4 G_3}
\end{equation}
implying that the entanglement curves are infinitely degenerate. However, when one considers fluctuations along with the orbifold symmetry, this infinite degeneracy is lifted, and only the $z=z_h, t=0$ curve represents the entanglement entropy between the two asymptotic defects. Using the HRT prescription \cite{Hubeny:2007xt, Wall:2012uf}, one finds that the entanglement entropy is given by
\begin{equation}
S_{\text{EE}}=\frac{\varphi_0+\varphi_h}{4G_3}.
\end{equation}
 
\subsection{Holographic complexity from KR braneworld}\label{mainresult}
In this section, we compute the holographic complexity using the `complexity$=$volume' proposal in the limit when the branes are nearly tensionless $s.t$ $\rho_{1,2}\approx \epsilon_{1,2}\ll 1$ and the fluctuations are even smaller than the brane locations ${\delta\varphi_{1,2}}\ll\rho_{1,2}$\footnote{In this limit $\mathcal{S}_{\text{dilaton}}$ goes as $\mathcal{O}(\epsilon^4)$. However, our result for complexity even holds exactly when we first consider orbifold symmetry with $\delta\varphi_2(y)=-\delta\varphi_1(y)$ for which $\mathcal{S}_{\text{dilaton}}$ vanishes and then take the tensionless limit to introduce some sort of locality in our setting.}. When the branes are nearly tensionless, the curvature scale on both branes matches the curvature scale of the bulk geometry, \textit{i.e.} $L_1\approx L_2\approx \ell$. The `complexity=volume' proposal dictates that the complexity of a dual quantum system is given by the maximal volume of the bulk codimension-1 hypersurface. We evaluate this volume casewise \textit{i.e.} first for rigid branes and then for the fluctuating branes. With two fixed AdS$_2$ branes, the bulk geometry is described by \ref{2nd},
\begin{eqnarray}
    ds^2&=&d\rho^2+\frac{\ell^2\cosh^2{(\rho/\ell)}}{u^2}\bigg[-f(u)d\tilde{t}^2+\frac{du^2}{f(u)}\bigg]\\
    &=&d\rho^2+\frac{\ell^2\cosh^2{(\rho/\ell)}}{z^2}\bigg[-h(z)dt^2+\frac{dz^2}{h(z)}\bigg]
\end{eqnarray}
where $h(z)$ is given by $h(z)=\big(1-\frac{z^2}{z_h^2}\big)$. The explicit coordinate transformation between $u$ and $z$ is,
\begin{equation}
    z=z_h\frac{u}{2 u_h-u},\,\,\,\,\,\,\,\,\,\,\,\,\tilde{t}=2t.
\end{equation}
We parametrize the volume surface by $t\equiv t(z)$. After doing this, we need to extremize the volume that is enclosed by these two rigid branes,
\begin{eqnarray}
    \mathcal{V}&=&\ell\int\int d\rho\,dz\frac{\cosh{(\rho/\ell)}}{z}\bigg[-h(z)t'^2+\frac{1}{h(z)}\bigg]^{1/2}\\
    &=&\ell^2\int\,dz\frac{\big(\sinh{(\rho_2/\ell)}-\sinh{(\rho_1/\ell)}\big)}{z}\bigg[-h(z)t'^2+\frac{1}{h(z)}\bigg]^{1/2}\\
    &\approx&\ell\int\,dz\frac{\big(\rho_2-\rho_1\big)}{z}\bigg[-h(z)t'^2+\frac{1}{h(z)}\bigg]^{1/2}\equiv \ell (\rho_2-\rho_1) \int dz \mathcal{L}\label{volempty}
\end{eqnarray}
The last line follows because we are working in the nearly tensionless limit, $\sinh{(\rho/\ell)} \approx \rho/\ell$. With this condition imposed, we now extremize the above volume functional in \ref{volempty}. The Lagrangian $\mathcal{L}$ in \ref{volempty} does not explicitly depends on time $t$; thus, we can find a conserved quantity which remains constant on the entire hyper-surface,
\begin{equation}
    \mathcal{E}=-\frac{\partial\mathcal{L}}{\partial {t'(z)}}\implies t_b=-\int_{0}^{z_t}\frac{\mathcal{E} z dz}{h(z) \sqrt{\mathcal{E}^2 z^2+h(z)}}\label{eqnconserve1}
\end{equation}
where $z_t$ is the turning point for the symmetric spacelike surface. At this point, the time derivative of the $z$ becomes zero. More precisely,
\begin{equation}
    \frac{1}{t'(z_t)}=0\implies z_t= \frac{z_h}{\sqrt{1-\mathcal{E}^2z_h^2}}
\end{equation} 
The maximal volume slice then follows by replacing \ref{eqnconserve1} into \ref{volempty},
\begin{equation}
    \mathcal{V}=2\ell^2 (\rho_2-\rho_1)\int_{0}^{z_t}\frac{dz}{z\sqrt{\mathcal{E}^2 z^2+h(z)}}\label{maxvolempty}
\end{equation}
Note that as boundary time $t_b$ approaches $\infty$, the conserved quantity $\mathcal{E}$ approaches to some critical value $\mathcal{E}_{\text{crit}}$. We find this critical value by extremizing $\mathcal{E}$ as the function of turning points $z_t$. This determines the critical value $\mathcal{E}_{\text{crit}}$ as a function of the horizon distance,
\begin{equation}
    \mathcal{E_\text{crit}}=\frac{1}{z_h}
\end{equation}
With the substitution of $\mathcal{E}_{\text{crit}}$ into \ref{maxvolempty} for the late times and taking the time derivative, we find the late time complexity growth,
\begin{equation}
     \frac{d\mathcal{C}}{dt_b}\bigg|_{t_b\to \infty}\approx\frac{2(\rho_2-\rho_1)}{G_3 z_h}\sim S_{\text{EE}} T\label{rigidcom}
\end{equation}
where $S_{\text{EE}}$ represents entanglement between two asymptotic defects when the branes are held fixed, and $T$ is the black hole temperature, $T=\frac{1}{2\pi z_h}$. The separation between two branes $(\rho_2-\rho_1)$ plays the role of the constant $\varphi_0$ in front of the topological term in JT gravity when both branes are fluctuating. Thus the growth rate in \ref{rigidcom} exactly matches the growth rate found in JT gravity in \ref{comJT}. Even though the entanglement entropy of the thermal state is constant over time, we find that the complexity of the state evolves with time, indicating the inherent thermal nature of the state.\par
We next proceed to compute the complexity while the branes are fluctuating. In that case, the bulk volume that is enclosed by these fluctuating branes is,
\begin{eqnarray}
    \mathcal{V}&=&\ell\int\int d\rho\,dz\frac{\cosh{(\rho/\ell)}}{z}\bigg[-h(z)t'^2+\frac{1}{h(z)}\bigg]^{1/2}\\
    &=&\ell^2\int \,dz\frac{\bigg(\sinh(\frac{\rho_2+\delta \varphi_2(y)}{\ell})-\sinh(\frac{\rho_1+\delta \varphi_1(y)}{\ell})\bigg)}{z}\bigg[-h(z)t'^2+\frac{1}{h(z)}\bigg]^{1/2}\\
    &=&\ell\int\,dz\bigg((\rho_2-\rho_1)+(\delta \varphi_2(y)-\delta \varphi_1(y))\bigg)\frac{1}{z}\bigg[-h(z)t'^2+\frac{1}{h(z)}\bigg]^{1/2}\\
    &=&\ell\int\,dz\bigg(\frac{\varphi_0+\varphi(z)}{z}\bigg)\bigg[-h(z)t'^2+\frac{1}{h(z)}\bigg]^{1/2}\\
    &=&\ell\int\,dz\frac{1}{z}\bigg(\varphi_0+\frac{\varphi_h {z_h}}{z}\bigg)\bigg[-h(z)t'^2+\frac{1}{h(z)}\bigg]^{1/2}\equiv \ell \int dz \mathcal{L}\label{volbtz}
\end{eqnarray}
We performed the radial integration as we went from the first to the second line in the above volume functional. From the second to the third line, we have used the fact that we are working in the nearly tensionless limit for the branes and thus $\sinh{x}\approx x$.
Note the new factor $\big(\varphi_0+\varphi(z)\big)$, which comes from the dimensional reduction of the bulk radial direction. Within this term, $\varphi(z)$ encapsulates the information of the brane fluctuations. This extra piece thus produces a significantly different result than complexity when the branes are rigid.
Furthermore, similar to before, the Lagrangian $\mathcal{L}$ in \ref{volbtz} still does not depend on time $t$ explicitly. Thus again we can find conserved quantity $\mathcal{E}$ for the volume surfaces,
\begin{equation}
    \mathcal{E}=-\frac{\partial\mathcal{L}}{\partial {t'(z)}}\implies t_b=-\int_{0}^{z_t}\frac{\mathcal{E} z^2 dz}{h(z) \sqrt{\mathcal{E}^2 z^4+(z \varphi_0+z_h\varphi_h)^2 h(z)}}\label{eqnconserve2}
\end{equation}
where $z_t$ is the turning point for the symmetric spacelike geodesics. By setting $\frac{1}{t'(z_t)}=0$, we find the energy for each volume hypersurface as a function of the turning point,
\begin{equation}
     \mathcal{E}=\frac{\sqrt{z_t^2-z_h^2} \left(\varphi_h z_h+\varphi_0z_t\right)}{z_h z_t^2}.
\end{equation}
We then find the maximal volume slice as a function of $\mathcal{E}$,
\begin{equation}
    \mathcal{V}=2\ell\int_{0}^{z_t}\frac{(\varphi_0 z+\varphi_h z_h)^2dz}{z^2\sqrt{\mathcal{E}^2 z^4+(z \varphi_0+z_h\varphi_h)^2 h(z)}}
\end{equation}
As we are interested in the behaviour of the complexity at late times, we find the value of $\mathcal{E}$ when $t_b\to\infty$. In this limit, $\mathcal{E}$ approaches to a critical value $\mathcal{E}_\text{crit}$. We determine $\mathcal{E}_\text{crit}$ by demanding that $\partial_{z_t}\mathcal{E}=0$ for the late times. This yields the critical value $\mathcal{E}_{\text{crit}}$ as a function of $\varphi_0$ and $\varphi_h$,
\begin{equation}
    \mathcal{E}_{crit}=\frac{\sqrt{2}(2\varphi_h^2+\varphi_0(\varphi_0+\sqrt{\varphi_0^2+8\varphi_h^2}))^{3/2}}{z_h(\varphi_0+\sqrt{\varphi_0^2+8\varphi_h^2})^2}
\end{equation}
By inserting the value of $\mathcal{E}_\text{crit}$ for the late times, we finally find the rate of growth of the complexity at the late times,
\begin{eqnarray}
   \frac{d\mathcal{C}_{\text{KR}}}{dt_b}\bigg|_{t_b\to \infty}&=&\frac{2\sqrt{2}(2\varphi_h^2+\varphi_0(\varphi_0+\sqrt{\varphi_0^2+8\varphi_h^2}))^{3/2}}{G_3 z_h(\varphi_0+\sqrt{\varphi_0^2+8\varphi_h^2})^2}\\
   &\approx&\frac{2\varphi_0}{G_3 z_h}\bigg(1+\frac{\varphi_h^2}{2\varphi_0^2}\bigg)+\mathcal{O}\big(\big({\varphi_h}/{\varphi_0}\big)^4\big)
\end{eqnarray}
where $C_{\text{KR}}$ denotes the complexity of the JT gravity that arises from the fluctuating KR branes.
The last line follows as we can Taylor expand the $rhs$ in ${\varphi_h}/{\varphi_0}\ll 1$. Thus, the complexity gets a sub-leading correction to the answer found in \cite{Brown:2018bms}.
This difference in complexities at the late times, $\Delta C=C_{\text{KR}}-C_{\text{JT}}$, grows as,
\begin{eqnarray}
    \frac{d\Delta C}{d t_b}\bigg|_{t_b\to\infty}=\frac{2\varphi_h^2}{G_3 z_h \varphi_0}+...\label{comdiff}
\end{eqnarray}
which is proportional to the $\varphi^2_h/\varphi_0$ while we held fixed the temperature. This subleading correction arises because of the brane fluctuations in bulk AdS and thus carries the signature of the fluctuations in the JT gravity complexity. In the limit when $G_3\to 0$ and $\varphi_h\to 0$ \footnote{The author in \cite{Geng:2022tfc} considers this limit to appropriately match the physics without any brane fluctuations, see section 8. }, the RHS of \ref{comdiff} vanishes, and we get the same expression for the complexity with rigid branes given by \ref{rigidcom}.\par
On a related note, it is well-known that JT gravity describes the near-horizon dynamics of near-extremal RN black holes in (3+1) dimension upon dimensional reduction. In that scenario, one identifies that $\varphi_0$ is related to the charge of the $(3+1)$ dimensional black hole. Using this \ref{comdiff} turns into,
\begin{equation}
    \frac{d\Delta C}{d t_b}\bigg|_{t_b\to\infty}=\frac{2\varphi_h^2}{G_3 z_h Q^2}
\end{equation}
where $Q$ is the charge of the near extremal RN black hole in $(3+1)$ dimension, defined as $\varphi_0=\frac {Q^2}{2}$. Schematically the complexity of the JT gravity goes as $\sim \# Q^2+\frac{\#}{Q^2}+...$.\\

\section{Discussions}\label{conc}
In this paper, we have studied the holographic complexity of Jackiw-Teitelboim (JT) arising from two fluctuating Karch-Randal branes in $(2+1)$ dimensions. These branes form a wedge in ambient AdS spacetime, and one can use the tools of wedge holography to study their low-energy effective dynamics. First, we consider rigid branes, which leads to the $2d$ Einstein-Hilbert gravity on the brane. By computing the maximal volume between these two branes, we find that at the late times, the complexity grows proportionally to the separation of the branes, which plays the role of $\varphi_0$ when the branes are fluctuating. This is somewhat expected as the complexity growth rate is proportional to entanglement entropy with a fixed temperature at the late times. As the entanglement entropy between the two asymptotic defects is proportional to the distance between two branes when they are held fixed; therefore it is correct to expect that we get the same sort of dependence for the complexity growth rate at the late times. Even though the entanglement entropy between the defects in the thermal state does not show any time dependence, the complexity grows with time, as expected for a thermal state of the boundary field theory. This stems from the basic expectation that although the entanglement is constant, the dual state still goes through nontrivial time evolution. Hence, although there is no change in the degree of entanglement, the complexity is supposed to capture the evolution in the state space. 

After that, we studied the holographic complexity of the thermal state when the bulk consists of two fluctuating AdS$_2$ branes. With the nearly tensionless branes, we find that the leading order term in the complexity grows proportionally with the $\varphi_0$ at the late times. Moreover, we find the first sub-leading correction is inversely proportional to $\varphi_0$. This is an entirely new fact. As reviewed in the main draft, the entanglement entropy for the fluctuating branes and the JT gravity theories match exactly. For the fluctuating branes in AdS$_3$, the HRT surface turns out to be non-degenerate and hence, unique. Unlike entanglement entropy, the subleading difference of complexity (equals volume conjecture) between the two theories suggests that the evolution of the two bulk theories is not completely equivalent in the state space. 
It is suggestive that complexity captures some nontrivial properties of the three-dimensional theory even after dimensional reduction, which does not seem to survive in a computation of entanglement entropy. It also strengthens the ``entanglement is not enough" proposal \cite{susskind2} in the sense that complexity can capture certain differences between the two theories in two dimensions, and hence the respective evolving states dual to the two theories, which entanglement entropy can not. Physically, we get this difference in the complexities because of the correct extremization of the volume functional. The authors in \cite{Anegawa:2023wrk} have recently suggested similar extremization of volume functional for de Sitter JT gravity; however, in terms of Weyl transformation of the 2$d$ intrinsic metric. They found an exact match for the de Sitter JT complexity with the dS$_3$ by appropriately choosing the warp factor $\Omega(r)=\varphi(r)^2$, which is exactly similar to our equation in \ref{volbtz}.
Nevertheless, we note down one further observation. While JT gravity is viewed as a low energy dynamics of a nearly extremal RN black hole in $(3+1)$ dimensions, the first subleading correction is found to be inversely proportional to the square of the total charge of the RN black hole. More concretely, the holographic complexity of the JT gravity grows as $\# Q^2+\frac{\#}{Q^2}+...$, where $Q$ is the charge of the RN black hole while the three-dimensional AdS curvature scale $\ell$ and dilaton at the horizon $\varphi_h$ are fixed. In the limit, $Q\to\infty$, only the first term survives, and all the sub-leading correction vanishes; thus, we recover the same result proposed in \cite{Brown:2018bms}.\par
There are some interesting future directions. An immediate direction would be considering the setup with orbifold symmetry and then evaluating the complexity of the JT gravity. In this paper, we only consider the tensionless branes; however, it would be worth checking that even with the orbifold symmetry of the bulk setup, we indeed get the same result for the complexity discussed in section \ref{mainresult}. Another possible future direction is to analyze the holographic complexity of the JT gravity by using the `complexity=action' proposal to see if this also gives similar results that we found using the `complexity=volume' proposal. This is a nontrivial exercise, as tackling the bulk action with fluctuations can be tricky. Finally, it is also an interesting question to make such a correspondence between the three-dimensional multi-boundary wormholes \cite{Caceres:2019giy} by performing a systematic dimensional reduction (which is also a topological theory defined on a timeslice of quotient AdS$_3$) and the JT gravity in the level of the action to compare the entanglement and complexity of the two theories along the lines of \cite{Bhattacharya:2020ymw, Bhattacharya:2020uun, Bhattacharya:2021dnd}. We hope to report in these directions soon. 
\section*{Acknowledgements}
 We thank Juan Pedraza for useful discussions and comments on the draft. A.B. $(1)$ is supported by the Institute of Eminence endowed postdoctoral fellowship offered by the Indian Institute of Science. A.B. $(2)$ is supported by Relevant Research Project grant (202011BRE03RP06633-BRNS) by the Board Of Research In Nuclear Sciences (BRNS), Department of atomic Energy, India and Mathematical Research Impact Centric Support Grant (MTR/2021/000490) by the Department of Science and Technology Science and Engineering Research Board (India). A.K.P is supported by the `Comunidad de Madrid' grant 2020-T1/TIC-20495 and the Spanish Research Agency (Agencia Estatal de Investigación) through the grants IFT Centro de Excelencia Severo Ochoa No. CEX2020-001007-S and PID2021-123017NB-I00, funded by MCIN/AEI/10.13039/501100011033 and by ERDF. 

\bibliographystyle{JHEP}
\bibliography{ComplexityinWedgeHolography}  

\end{document}